\documentclass[aps,prl,reprint]{revtex4-2}

\usepackage{amsmath, amssymb}
\usepackage{graphicx}
\usepackage{dcolumn} 
\usepackage{bm}      
\usepackage{booktabs}
\usepackage{physics}
\usepackage{tikz}
\usepackage{tikz-3dplot}
\usetikzlibrary{decorations.pathmorphing, arrows.meta, calc, shapes, shadows, patterns}
\usepackage{units}

\newcommand{\ie}{\textit{i.e.,}\ }

\newcommand{\etal}{\textit{et al.}}

\emergencystretch=\maxdimen
\begin{document}
	
	\title{Detecting Solenoidal Plasma Turbulence via Laser Polarization Rotation}
	
	\author{Kenan Qu}
	\affiliation{Department of Astrophysical Sciences, Princeton University,  Princeton, New Jersey 08544, USA \looseness=-1 }  
	\author{Nathaniel J. Fisch}
	\affiliation{Department of Astrophysical Sciences, Princeton University,  Princeton, New Jersey 08544, USA \looseness=-1 }  
	
	\date{\today}
	
	\begin{abstract}
	Recent theoretical studies suggest that solenoidal turbulence can significantly enhance fusion reactivity, yet no standard diagnostic exists to directly measure these solenoidal flows in high-energy-density plasmas, nor to distinguish between solenoidal and compressional turbulence. 
		We propose a method that directly diagnoses the energy and spatial structure of this rotational turbulence using the cross-polarization scattering of a probe laser. By coupling to the plasma vorticity, the scattering generates a cross-polarized signal proportional to the turbulent vorticity, effectively acting as a calorimeter for shear flows. We identify a diffractive scattering signature analogous to  ``Debye-Scherrer ring'' that reveals the eddy size distribution. We show that this technique is applicable to National Ignition Facility (NIF) implosion conditions and  other high-energy-density scenarios.
	\end{abstract}
	
	\maketitle
	
	\textit{Introduction---}	
	Turbulence plays a central role in the dynamics of high-energy-density (HED) plasmas, particularly in the context of Inertial Confinement Fusion (ICF). Physically, turbulent flows decompose into two distinct components: compressional modes ($\nabla \cdot \mathbf{u} \neq 0$, where $\mathbf{u}$ is the fluid velocity), associated with density fluctuations, and solenoidal or rotational modes ($\nabla \times \mathbf{u} \neq 0$), which manifest as shear flows and vortices \cite{Lindl1995, Remington2006}.
	Historically, experimental diagnostics have focused heavily on the compressional component. Tools such as Thomson scattering, interferometry, and radiometry are highly sensitive to electron density fluctuations, making the measurement of compressional turbulence relatively straightforward \cite{Froula2011, Glenzer2009}. Various implementations of these techniques, such as Doppler backscattering and cross-polarization scattering (for magnetic fluctuations), have been successfully deployed in magnetic fusion devices to characterize density turbulence \cite{Rhodes2012, Rhodes2014, Barada2016, Shi2022, Zhang_2025, Zou1995, Lehner1989, Maron2020}.
	
	However, the recently proposed ``shear flow reactivity enhancement effect'' suggests that solenoidal turbulence can significantly boost fusion reactivity beyond levels expected from thermalization alone \cite{Fetsch2025a, Fetsch2025b}. In this regime, fast ions with long mean free paths sample differing bulk velocities across turbulent eddies, effectively increasing the center-of-mass collision energy. This finding challenges the conventional view that imploding plasmas must fully thermalize to maximize yield and suggests that embracing, rather than avoiding, shear flow turbulence could lower the ignition threshold.
	
	In view of this possibly profound paradigmatic shift, a major experimental gap is exposed: there are no established, non-intrusive methods to directly detect and quantify these shear flows in dense HED plasmas \cite{Nilson2025, Hammel2010}. Because solenoidal motions are incompressible to first order in the turbulent energy, they do not generate the density fluctuations required for standard scattering techniques. Without a diagnostic capable of isolating the rotational component of turbulence, experimental verification of these shear-flow theories remains elusive.
	
	Polarimetry offers a promising path forward. As a diagnostic technique sensitive to the polarization of the electromagnetic field, it is inherently well-suited for detecting transverse fluctuations in the plasma medium, such as shear flows. However, the application of polarimetry to unmagnetized turbulence has been historically overlooked. In conventional plasma diagnostics, polarization rotation is almost exclusively associated with the presence of a background magnetic field (known as Faraday rotation \cite{Hutchinson2002, Donne1999, Segre1999}), leading to the widespread assumption that polarimetry is ineffective in unmagnetized HED environments.
	
	This assumption stems from a reliance on magnetic gyrotropy. It is often assumed that unmagnetized plasmas, lacking this symmetry breaking, would not exhibit such polarization effects. However, recent work by Gu\'{e}roult and Langlois \cite{Langlois2024, Gueroult2023} has demonstrated that a rotating unmagnetized plasma possesses an effective gyrotropy due to inertial forces, \textit{i.e.}, the Coriolis effect. This polarization drag effect allows us to build a diagnostic tool specifically sensitive to plasma vorticity.
	
	While previous work established the fundamental mechanism for coherent, rigid-body rotation of a bulk plasma, the implications for random, multiscale turbulence have remained unexplored. Here, we propose utilizing this mechanism as a targeted diagnostic for solenoidal turbulence. We demonstrate that plasma fluid \textit{vorticity} acts as a source term for electromagnetic scattering, thereby modifying the polarization state of a probe beam.
	Unlike the coherent rotation described by Gu\'{e}roult \etal, we describe an effect where randomly distributed turbulent eddies cause an incoherent accumulation of polarization rotation, manifesting as a measurable rms spreading of the polarization angle. This allows the energy of the rotational velocity field to be inferred directly from the cross-polarized scattered light, independent of density fluctuations. Furthermore, we exploit the diffractive nature of this scattering, specifically the formation of Debye-Scherrer rings, to determine the spatial scale of the turbulence from the ring's opening angle.
	
	Thus, by simultaneously retrieving the two scalar quantities of solenoidal turbulence that govern the shear flow reactivity enhancement effect---the  energy magnitude and its characteristic scale---this methodology enables the key 3D tomographic reconstruction of the turbulence.

	\textit{Mechanism of Polarization Rotation---}	\label{sec:mechanism}	
	The polarization rotation can be seen by considering the different phase velocities of circular polarization components in the rotating frame. A linearly polarized wave can be decomposed into right-hand and left-hand circular components. In a frame rotating at $\omega_{\text{rot}}$, each component experiences a rotational Doppler frequency shift $\Delta\omega = \pm\omega_{\text{rot}}$, leading to a phase velocity difference $\pm(\omega_p^2 / \omega_L^2)(\omega_{\text{rot}}t/2)$, where $\omega_p$ is the plasma frequency and $\omega_L$ is the laser frequency.
	This differential phase accumulation results in a polarization rotation angle
	\begin{equation}
		\psi =\frac{L}{c\eta } \frac{n_e}{n_c} \omega_{\text{rot}},
		\label{eq:psi_coh}
	\end{equation}
	where $L$ is the interaction length, $c$ is the speed of light, $n_e$ is the electron density, $n_c = m_e\epsilon_0 \omega_L^2/e^2$ is the critical plasma density, $\eta = \sqrt{1-n_e/n_c}$ is the non-relativistic refractive index, $m_e$ is the electron rest mass, $\epsilon_0$ is the vacuum permittivity, and $e$ is the natural charge. 
	The scattering depends on three factors: the fluid vorticity $\Omega$, the plasma density $n_e$, and the refractive index $\eta$. Critically, the scattering is amplified for plasma near the critical density ($\eta\to 0$). The reduced laser group velocity increases the interaction time as the beam traverses the plasma turbulence.
	
	The effect of a rotating rigid medium dragging the plane of polarization of light passing through it was investigated under the term ``Fresnel drag'' by Fermi~\cite{Fermi1923}, and later by Jones~\cite{Jones1976} and Player~\cite{Player1976}. The theory was recently extended to plasmas by Gu\'{e}roult \etal~\cite{Langlois2024} by assuming that the dielectric conductivity tensor transforms rigidly with plasma rotation. It can be demonstrated that this polarization rotation angle is robust at temperatures in a fusion environment.
	
	\textit{Depolarization due to Incoherent Scattering---}	
	Equation (\ref{eq:psi_coh}) applies if the entire plasma rotates as a rigid body. However, in realistic plasma turbulence, the vorticity field $\boldsymbol{\omega}_{\text{rot}}(\mathbf{x})$ is not a single coherent structure but a collection of random eddies. This process is most accurately described as a random rotation of the polarization angle. As the laser propagates through the plasma, it encounters a sequence of uncorrelated vortices, each imparting a small, random polarization rotation $\delta \psi$.
	Physically, this manifests as depolarization: a linearly polarized beam broadens into an elliptical distribution with a variance proportional to the interaction length, as illustrated in Fig.~\ref{fig:random_walk}.
	
	\begin{figure}[t]
		\centering
		\begin{tikzpicture}[>=latex, scale=0.9]
			\foreach \x/\tilt/\mag/\arrowtype in {
				1/15/0.8/\circlearrowleft, 
				2.5/-20/1.3/\circlearrowright, 
				4.0/45/1.8/\circlearrowleft, 
				5.5/-10/1.0/\circlearrowright} {
				
				\draw[fill=cyan!10, draw=cyan!50, dashed, rotate around={\tilt:(\x,0)}] (\x,0) ellipse (0.5 and 0.25);
				\node[font=\footnotesize, scale=\mag, rotate=\tilt] at (\x,0) {$\arrowtype$};
			}
			
			\draw[->] (0,0) -- (7,0) node[right] {$z$};
			\draw[->] (0,-1.2) -- (0,1.2) node[above] {$\psi$};
			\draw[red, thick] (0,0) -- (1, 0.1) -- (2.5, -0.2) -- (4, 0.3) -- (5.5, 0.1) -- (7, 0.4);
			\draw[orange, dashed, thick] plot[domain=0:7] (\x, {0.15*sqrt(\x)});
			\node[orange, right, font=\footnotesize] at (7, 0.4) {$\psi_{\text{rms}} \propto \sqrt{L}$};
			\draw[orange, dashed, thick] plot[domain=0:7] (\x, {-0.15*sqrt(\x)});
			
		\end{tikzpicture}
		\caption{\label{fig:random_walk} Random walk of the polarization angle $\psi$ as the laser beam traverses multiple uncorrelated turbulent eddies, leading to an rms broadening $\psi_{\text{rms}}$.}
	\end{figure}
	
	If the size of each eddy is $l_{\text{eddy}}$, the accumulated polarization change after interacting with $N=L/l_{\text{eddy}}$ eddies follows a random walk
	\begin{equation}
		\langle \Delta \psi^2 \rangle = \sum_{i=1}^N \langle \delta \psi_i^2 \rangle =  \left(\frac{1}{c\eta } \frac{n_e}{n_c}\omega_{\text{rot}} \right)^2 L{l_{\text{eddy}}} .
	\end{equation}
	Consequently, the rms rotation angle $\psi_{\text{rms}} = \sqrt{\langle \Delta \psi^2 \rangle}$ scales with the square root of the interaction length.
	
	\textit{Diagnostic of Turbulent Energy---}
	We can invert these relations to serve as a diagnostic for the kinetic energy of the plasma turbulence. The turbulent kinetic energy density $W_{\text{turb}}$ is defined as $W_{\text{turb}} = \frac{1}{2} m_e n_e \langle u^2 \rangle$. Assuming isotropic turbulence, the vorticity magnitude is defined by the velocity shear at the eddy scale, $\omega_{\text{rot}} \approx u_{\text{rms}}/l_{\text{eddy}}$ (assuming eddies have comparable transverse and longitudinal dimensions).
	We then find a scaling relation dependent on the turbulence parameters
	\begin{equation}
		\psi_{\text{rms}} = \frac{\sqrt{2}}{c\eta} \left( \frac{n_e}{n_c} \right) \sqrt{\frac{L}{l_{\text{eddy}}}} \sqrt{\frac{W_{\text{turb}}}{m_e n_e}}.
		\label{eq:psi_rms}
	\end{equation}
	The polarization rotation $\psi$ gives rise to a cross-polarized electric field $E_\perp = E_{\text{in}} \sin\psi \approx E_{\text{in}} \psi$, implying that the measured power fraction scales quadratically with the angle, $P_\perp/P_{\text{in}} \approx \psi_{\text{rms}}^2$.
	In an experiment, one typically measures the total power or energy of the cross-polarized light ($P_\perp$) relative to the incident power ($P_{\text{in}}$)
	\begin{equation}\label{eq:power_ratio}
		\frac{P_\perp}{P_{\text{in}}} = \frac{2 W_{\text{turb}}}{m_e c^2 \eta^2}  \left( \frac{n_e}{n_c^2} \right)  \left( \frac{L}{l_{\text{eddy}}}\right).
	\end{equation}
	The scattered power ratio is linearly proportional to the total turbulent kinetic energy. Thus, the cross-polarized detector acts as a {calorimeter} for plasma turbulence.
	To obtain the turbulence energy density $W_{\text{turb}}$, one requires the plasma density $n_e$ (measurable via standard Thomson scattering or interferometry) and the eddy size $l_{\text{eddy}}$. In the following section, we describe how $l_{\text{eddy}}$ can be measured directly from the scattering geometry.
	
	\textit{Relation between Scattering Angle and Eddy Size---}
	\label{sec:signatures}	
	The frequency-conserving scattering process due to random plasma turbulence can be treated as the elastic diffraction of a coherent input. While the total cross-polarized power is a calorimetric measure of the energy, the spatial structure of the scattering encodes the geometric properties of the turbulence.

	The collective scattering of eddies with characteristic size $l_{\text{eddy}}$ produces a cone-shaped diffraction pattern. Eddies with a distribution of sizes will form a distribution of rings (or a diffuse halo), where each radial angle $\vartheta$ corresponds to a specific eddy size $l_{\text{eddy}}$ in the spectrum.
	The scattered field is proportional to the Fourier transform of the vorticity field ${\boldsymbol{\Omega}}(\mathbf{q})$ evaluated at the scattering vector $\mathbf{q} = \mathbf{k}_s - \mathbf{k}_L$, defined by the scattered ($\mathbf{k}_s$) and incident ($\mathbf{k}_L$) wave vectors. For isotropic turbulence characterized by a dominant eddy scale $l_{\text{eddy}}$, the vorticity power spectrum $S_\Omega(\mathbf{q}) = \langle |{\boldsymbol{\Omega}}(\mathbf{q})|^2 \rangle$ is concentrated in a spherical shell in $k$-space with radius $q_0 = 2\pi/l_{\text{eddy}}$.
	Elastic scattering conserves frequency, \ie $|\mathbf{k}_s| = |\mathbf{k}_L| = k_L$. Geometrically, this constrains the scattering vector $\mathbf{q}$ to lie on the intersection of the Ewald sphere (radius $k_L$) and the turbulence spectral shell. This intersection defines a cone of angle $\vartheta$ relative to the $\mathbf{k}_L$ direction
	\begin{equation} \label{eq:q0}
		q = 2 k_L \sin(\vartheta/2) = q_0.
	\end{equation}
	For small scattering angles, this simplifies to the Bragg-like condition
	\begin{equation}
		\vartheta \approx \frac{q_0}{k_L} = \frac{\lambda_{L}}{l_{\text{eddy}}},
	\end{equation}
	where $\lambda_L$ is the laser wavelength. For isotropic turbulence with random eddy orientations, the scattering is azimuthally symmetric around the laser axis. To isolate this signal, a cross-polarization filter is placed in the beam path, extinguishing the coherent unscattered light while transmitting the rotationally scattered components.

	If the turbulence possesses a single dominant length scale, the scattered light forms a distinct ring on a downstream detector, analogous to the Debye-Scherrer rings observed in X-ray powder diffraction \cite{Warren1990}. For vortices with totally random sizes, the rings would defuse into a filled disk. However, the distinct ring structure remains observable even for eddies following a standard Kolmogorov cascade or  any generalized turbulent energy cascade with the scaling $q^{-p}$ for $p<2$ (see End Matter). Furthermore, the signal at a specific radius (angle $\vartheta$) still corresponds to a specific eddy size $l_{\text{eddy}} \approx \lambda_L/\vartheta$, and the angular intensity profile $I(\vartheta)$ allows for the reconstruction of the turbulence size distribution $S(l_{\text{eddy}})\propto \vartheta^2  I(\vartheta)$. Thus, by measuring the radial profile of the Debye-Scherrer pattern in the cross-polarized channel, one can directly measure the power spectrum of the turbulent eddies.

\begin{figure}[htbp]
    \centering
    \includegraphics[width=\columnwidth]{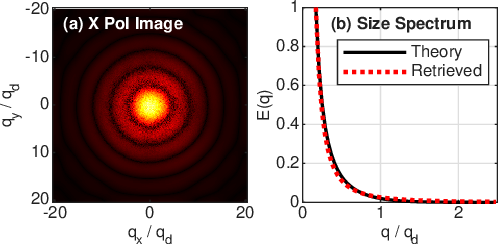} 
    \caption{(a) Reconstructed power spectrum $S_{\Omega}(\mathbf{q})$ of the plasma vorticity field using the incident X ray beam as a self-reference. (b) Normalized eddy size spectrum. Here, $n_e=0.99n_c$ and $\delta n_e\leq 0.004n_c$. }
    \label{fig:simu}
\end{figure}

    Remarkably, this polarimetric diagnostic remains robust against macroscopic density fluctuations, even when probing near-critical plasmas where refractive sensitivity is highly amplified. While density gradients typically act as stochastic lenses that severely distort the propagating wave by imprinting a spurious spatial phase shift $\Phi \propto \int \delta n_e \, dz$, the incident X ray can serve as a ``self-referencing" adaptive optics principle to eliminate this smearing, as shown in Fig~\ref{fig:simu} (see End Matter for more details). Because the intense driving beam and the cross-polarized turbulence-scattered signal copropagate simultaneously through the exact same macroscopic density landscape, they acquire the identical refractive phase modulation $e^{i\Phi}$. By measuring the distorted intensity profile of the bright transmitted beam on the detector, we obtain a direct, in-situ measurement of the system's macroscopic Point Spread Function, $\text{PSF} \propto |\mathcal{F}\{E_\mathrm{in} e^{i\Phi}\}|^2$. The diffuse cross-polarized image, which mathematically represents a convolution of the turbulence diffractive signature with this exact PSF, can then be computationally deconvolved, effectively stripping away the stochastic refractive noise to fully recover the underlying true eddy size spectrum.

	It is important to decouple the information contained in the ring's geometry from its brightness. The ring radius (angle $\vartheta$) is determined by the eddy size $l_{\text{eddy}}$ via diffraction; smaller eddies scatter light to wider angles. Measuring the ring radius provides the spatial scale of the turbulence $l_{\text{eddy}}$, which can then be substituted back into Eq.~(\ref{eq:q0}) to isolate the turbulent energy density.
	The ring brightness, on the other hand, is determined by the vorticity magnitude and plasma density. A more energetic vortex makes the ring brighter but does not change its diameter.

    The spectral width of the diffraction ring provides a measurement of the turbulence kinematics that is complementary to the polarization rotation method. While the polarization rotation (manifesting as the total cross-polarized power) yields a calorimetric measure of the turbulent energy ($P_\perp \propto n_e^2 u_{\text{rms}}^2$), the spectral linewidth independently encodes the velocity distribution. Since the scattering vector magnitude $q_0$ is fixed by the ring angle, the spectral lineshape is determined by the Doppler broadening from the random eddy velocities, scaling as $\Delta \omega_{\text{ring}} \approx q_0 u_{\text{rms}}$. Thus, spectrally resolving the ring allows one to extract the vorticity magnitude $|\boldsymbol{\Omega}|$ and verify the fluid velocity scale separate from the signal brightness.
    
    This spectral signature is also useful for isolating the turbulence from the thermal background. Although the collected light is an incoherent sum of scattering from many eddies, resulting in a broadened peak, this peak remains distinct because of the separation of velocity scales. In typical HED conditions, the electron thermal velocity ($v_{th}$) exceeds the fluid velocity ($u_{\text{rms}}$) by orders of magnitude ($v_{th} \gg u_{\text{rms}}$). Consequently, the thermal electrons produce a wide, diffuse spectral pedestal ($\Delta \omega_{\text{th}} \sim k v_{th}$), whereas the vorticity scattering appears as a comparatively narrow peak ($\Delta \omega_{\text{ring}} \sim k u_{\text{rms}}$) superimposed on top. However, if the fluid velocity were to approach the thermal velocity ($u_{\text{rms}} \to v_{th}$), the spectral broadening of the turbulence would become comparable to the thermal width. In this limit, the distinctive narrow peak would widen and vanish into the thermal noise floor, making the diagnostic ineffective.
	
	\textit{Numerical Examples and Experimental Feasibility---}
	\label{sec:numerical}	
	To assess the practicality of this diagnostic, we evaluate the expected signal strength for three distinct experimental regimes in HED physics. The primary metrics are the rms polarization rotation angle $\psi_{\text{rms}}$, the cross-polarized intensity ratio $P_{\perp}/P_{\text{in}}$, and the Debye-Scherrer ring angle $\vartheta$. The results are summarized in Table~\ref{tab:estimates}. 
	
	\begin{table}[h]
		\centering
		\caption{Summary of Feasibility Estimates}
		\label{tab:estimates}
		\resizebox{\columnwidth}{!}{%
			\begin{tabular}{lccc}
				\toprule
				Parameter & Case A & Case B & Case C (Far-IR) \\
				& (NIF) & (Ablation) & (Gas Jet) \\ \midrule
				Probe $\lambda$ & 30 nm & 0.33 $\mu$m & 33 $\mu$m \\
				Density $n_e$ & $10^{24}$ cm$^{-3}$ & $10^{22}$ cm$^{-3}$ & $10^{18}$ cm$^{-3}$ \\
				Refr. Index $\eta$ & 0.1 & 0.14 & 0.1 \\
				Eddy Size $l_{\text{eddy}}$ & $1\, \mu\text{m}$ & $7\, \mu\text{m}$ & $2$ mm \\
				Interactions $N$ & 50 & 57 & 1 \\
				Flow Vel. $u$ & 300 km/s & 6 km/s & 2 km/s \\
				Pol. Rot. $\delta \psi$ & 10 mrad & 0.15 mrad & 0.07 mrad \\ \midrule
				RMS Rotation & 70 mrad & 1.1 mrad & 0.07 mrad \\
				Intensity Ratio & $5 \times 10^{-3}$ & $10^{-6}$ & $5 \times 10^{-9}$ \\
				Ring Angle $\vartheta$ & 30 mrad & 47 mrad & 17 mrad \\
				\bottomrule
			\end{tabular}
		}
	\end{table}
	
	\textit{Case A: NIF Implosion.} This scenario represents the most promising application. In ICF experiments at the National Ignition Facility (NIF), the mix layer between the shell and fuel is extremely dense ($n_e \sim 10^{24}$ cm$^{-3}$) and turbulent \cite{Hammel2010}. We assume a soft X-ray probe ($\lambda \approx 30$ nm) chosen to be close to the plasma frequency of the mix layer ($n_e/n_c \approx 0.99$).
	The interaction in this regime is dominated by the proximity to the critical density. As $n_e \to n_c$, the refractive index $\eta$ drops to $\sim 0.1$. The slow group velocity dramatically increases the interaction time and effective coupling efficiency between the probe and the turbulence, amplifying the scattered signal significantly compared to underdense interactions.
	
	A critical requirement for this diagnostic is a probe beam with a high degree of initial linear polarization. For the X-ray regime, several sources can provide the necessary polarization purity. Synchrotron radiation and X-ray Free Electron Lasers (XFEL) are naturally highly polarized with extinction ratios below $10^{-5}$ \cite{Schulze2022, Lutman2016}. Laser-produced High Harmonic Generation (HHG) sources are also ideal as they inherit the linear polarization of the drive laser with high fidelity, achieving extinction ratios better than $10^{-4}$. Alternatively, one can use crystal optics based on Bragg reflection at $45^\circ$ to purify the polarization of thermal X-ray sources \cite{Marx2013}.
	
	Regarding detection, the incoherent accumulation over $N \approx 50$ interactions yields a rotation $\psi_{\text{rms}} \sim 70$ mrad. This corresponds to a cross-polarized intensity ratio $P_\perp/P_{\text{in}} \approx 5 \times 10^{-3}$. Assuming an input X-ray pulse with $10^{12}$ photons, the cross-polarized signal would contain $\sim 5 \times 10^9$ photons, a signal easily distinguishable from noise. Even if the signal is weakened by plasma lensing effects, it would remain above the extinction floor of input polarization.
	The expected ring angle is $\vartheta \approx 30\,\text{nm} / 1\,\mu\text{m} \approx 0.03$ rad ($1.7^\circ$). This angle allows the scattered signal to be spatially separated from the main beam while remaining within the collection cone of standard area detectors.
	
	\textit{Case B: Turbulent Laser-Produced Plasma.} This case corresponds to recent experiments studying turbulence in laser-produced plasmas \cite{Rigon2021}, where a shock driven into a low-density foam target generates a turbulent mixing zone. The characteristic plasma parameters observed are an electron density $n_e \approx 10^{22}$ cm$^{-3}$ and a fluid velocity $u \approx 6$ km/s.
	The characteristic turbulence scale length, identified as the spectral knee, is $l_{\text{eddy}} \approx 7\,\mu$m, and the plasma extension is $L \approx 400\,\mu$m. To probe this high-density plasma, we propose using a UV probe ($\lambda = 0.33\,\mu$m, e.g., a frequency-tripled Nd:YAG laser or excimer laser).
	The estimated rms rotation is $\psi_{\text{rms}} \approx 1.1$ mrad, corresponding to an intensity ratio of $P_\perp/P_{\text{in}} \sim 1.2 \times 10^{-6}$. While weaker than the resonant NIF case due to the lower flow velocity, this signal is well within the range of standard high-extinction polarimetry, making detection feasible. The diffractive ring angle $\vartheta \approx 0.33\,\mu\text{m} / 7\,\mu\text{m} \approx 47$ mrad ($2.7^\circ$) is well-suited for detection.
	
	\textit{Case C: Gas Jet.} This scenario is typical of laser wakefield acceleration experiments using low-density gas targets. Using a standard infrared probe ($\lambda = 0.8 \mu$m), the predicted signal is $\psi \approx 50\,\mu$rad, corresponding to an intensity ratio of $P_\perp/P_{\text{in}} \sim 10^{-9}$. This level is practically unmeasurable due to photon statistics and background noise.
	However, by utilizing a far-IR probe (e.g., $\lambda = 33\,\mu$m), we can tune the diagnostic to the critical density of the tenuous plasma ($n_c \approx 10^{18}$ cm$^{-3}$). With this resonant enhancement ($\eta \approx 0.1$), the signal is dramatically boosted. The rms rotation angle increases to $\psi_{\text{rms}} \approx 0.07$ mrad, corresponding to an intensity ratio of $P_\perp/P_{\text{in}} \approx 5 \times 10^{-9}$. This renders the measurement feasible, demonstrating the versatility of the method across density regimes. The ring angle would be $\vartheta \approx 33\,\mu\text{m} / 2\,\text{mm} \approx 17$ mrad ($1^\circ$).
	
	\textit{Discussion and Conclusion---}
	\label{sec:discussion}
	We showed how laser polarization rotation can diagnose plasma turbulence, specifically targeting the vortical component critical to fusion yield. The method is particularly useful because it scales with the energy of the turbulence rather than merely the density fluctuations. By measuring the total power of the cross-polarized scattered light, one obtains a calorimetric measurement of the turbulent kinetic energy.
	In HED regimes, specifically near-critical plasmas like those in NIF implosions, the effect is resonantly enhanced, offering a unique window into the dynamics of hydrodynamic instabilities that are otherwise invisible to standard diagnostics. Critically, this diagnostic bridges the gap between turbulence theory and fusion performance.
	
	The extracted values of turbulent kinetic energy $W_{\text{turb}}$ and eddy size $l_{\text{eddy}}$ are the precise inputs required to evaluate the ``shear flow reactivity enhancement'' factor predicted by Fetsch and Fisch \cite{Fetsch2025a}. By quantifying $W_{\text{turb}}$, one can estimate the effective temperature increase experienced by reacting ions, thereby directly determining whether the turbulence is acting as a deleterious energy sink or a beneficial reactivity booster. This capability could be pivotal for optimizing ignition designs that leverage, rather than suppress, shear flows.
	
    
    A potential complication is the presence of spontaneous magnetic fields, which are common in laser-produced plasmas through, e.g., the Biermann battery effect \cite{Stamper1971}. These fields also rotate the polarization plane via the classical Faraday effect. However, the magnetic and mechanical rotations are distinguishable. First, the macroscopic magnetic field can be independently measured by other means, such as detecting the deflection of a charged particle beam (e.g., proton radiography), allowing its specific Faraday contribution to be calculated and isolated. Second, magnetic fields and kinetic turbulent eddies typically exhibit fundamentally different dissipation lifetimes. By taking time-resolved measurements across multiple shots, one can temporally decouple the transient vortical scattering from the longer-lived magnetic structures.

	Despite these caveats, it remains that this methodology offers a  tool uniquely suited to diagnose directly solenoidal turbulence. 	Furthermore, by employing multiple probe beams at different angles and polarization states, it is theoretically possible to reconstruct the three-dimensional vorticity vector field $\boldsymbol{\Omega}(\mathbf{x})$ using vector tomography techniques \cite{Cormack1964, Granetz1988, Prince1995}, analogous to methods used in medical imaging and tokamak SXR diagnostics.

	\begin{acknowledgments}
		This work was supported by  NNSA Grant No. DE-NA0004167 and NSF Grant No. PHY-2308829. The authors gratefully acknowledge valuable discussions with Henry Fetsch.
	\end{acknowledgments}  
	
	\bibliography{references.bbl}

\pagebreak 
\setcounter{equation}{0}
\renewcommand{\theequation}{S\arabic{equation}}
\setcounter{figure}{0} 
\renewcommand{\thefigure}{S\arabic{figure}} 
\setcounter{secnumdepth}{2}

\appendix

\section{Diffractive Imaging Model and Self-Referencing}

To model the diffractive imaging of plasma eddies, we extend our framework to account for a probe beam of finite transverse width and the simultaneous stochastic refraction caused by background density perturbations. We operate in the paraxial limit, assuming the X-ray propagates predominantly along the $z$-axis with transverse scattering taking place in the $\mathbf{r}_\perp = (x,y)$ plane. We explicitly account for a background plasma density $n_0$ that may be close to the critical density $n_c$.

Consider a linearly polarized incident X-ray beam (polarized along $\hat{x}$) with a finite transverse envelope $E_\text{in}(\mathbf{r}_\perp)$, where the beam waist $W$ is larger than the characteristic eddy size $l_{eddy}$. 
The background plasma has a refractive index $\eta_0 = \sqrt{1 - n_0/n_c}$. As the beam propagates through the plasma of length $L$, macroscopic density fluctuations $\delta n(\mathbf{r}_\perp, z)$ modify the local refractive index. Expanding the refractive index for a near-critical plasma ($n_0 \lesssim n_c$), we obtain
\begin{equation}
    \eta(\mathbf{r}) = \sqrt{1 - \frac{n_0 + \delta n}{n_c}} 
    \approx \eta_0 - \frac{1}{2\eta_0}\frac{\delta n}{n_c}.
\end{equation}
Assuming the density gradient is small within a wavelength, the cumulative refractive phase shift $\Phi(\mathbf{r}_\perp)$ accrued along the propagation axis can be treated via the eikonal approximation
\begin{equation}
    \Phi(\mathbf{r}_\perp) = -\frac{k_L}{2\eta_0 n_c} \int_0^L \delta n(\mathbf{r}_\perp, z') dz'.
\end{equation}


Because the scattering due to polarization drag is forward-directed and the diffraction angles are small ($\theta \ll 1$), we can project the 3D scattering process onto an effective 2D exit plane. The total cross-polarized source field $E_y$ generated at the exit plane $z=L$ is the superposition of the fields generated by all eddies, multiplied by the beam envelope and the refractive phase
\begin{equation}
    E_{y}(\mathbf{r}_\perp) = i \frac{\kappa}{\eta_0} \left[ \sum_{n=1}^N \omega(\mathbf{r}_\perp - \mathbf{\rho}_n) \right] E_\text{in}(\mathbf{r}_\perp) e^{i\Phi(\mathbf{r}_\perp)},
\end{equation}
where $\kappa/\eta_0$ represents the enhanced coupling constant for the polarization rotation in a near-critical medium, and $\omega(\mathbf{r}_\perp)$ is the characteristic 2D transverse profile of the vorticity inside a single eddy. 

It might intuitively seem that the cross-polarized light generated by an eddy at $z_n$ should only suffer refractive smearing from $z_n$ to the exit plane $L$. However, we must account for the phase continuity of the incident beam. By the time the incident wave reaches the eddy at $z_n$, it has already accumulated a pre-scattering phase distortion. The total phase at the exit plane is the linear sum of the pre-scattering and post-scattering phases, which evaluates exactly to the full integral from $0$ to $L$. 

To find the resulting diffraction image on a detector placed at a far-field distance $D$, we take the Fraunhofer diffraction integral (or 2D spatial Fourier transform) of the exit field. Let $\mathbf{q}$ be the transverse scattering wavevector, where $|\mathbf{q}| = k_L \sin\theta \approx k_L \theta$.
The cross-polarized field at the detector plane is
\begin{align}
    \tilde{E}_y(\mathbf{q}) &= \int d^2\mathbf{r}_\perp E_{y}(\mathbf{r}_\perp) e^{-i \mathbf{q} \cdot \mathbf{r}_\perp} \nonumber \\
    &= \tilde{V}(\mathbf{q}) \ast \tilde{E}_\text{in}(\mathbf{q}) \ast \tilde{P}_{ref}(\mathbf{q}),
\end{align}
where $\ast$ denotes the 2D convolution, and $\tilde{V}(\mathbf{q}) = \tilde{\omega}(\mathbf{q}) \sum_{n=1}^N e^{-i \mathbf{q} \cdot \mathbf{\rho}_n}$ denotes the ideal scattering geometry. The summation term generates the rapid, high-frequency laser speckle pattern typical of random multi-scatterer interference. 
The final intensity measured by the polarimetric calorimeter is thus 
\begin{equation} \label{eq:Iperp}
    \langle I(\mathbf{q}) \rangle \propto \frac{N}{\eta_0^2} \left| \tilde{\omega}(\mathbf{q}) \ast \tilde{E}_\text{in}(\mathbf{q}) \ast \tilde{P}_{ref}(\mathbf{q}) \right|^2.
\end{equation}

The next two terms $\tilde{E}_\text{in}(\mathbf{q})$ and $\tilde{P}_{ref}(\mathbf{q}) = \mathcal{F}\{e^{i\Phi(\mathbf{r}_\perp)}\}$ describe the Fourier transform of the incident beam and the refractive steering due to density perturbations, respectively. Crucially, this multiplication could be directly recovered by detecting the diffraction of the incident beam.  
Because the incident beam propagates through the identical macroscopic density gradients as the cross-polarized scattered emission, it provides a simultaneous, exact measurement of the system's instantaneous Point Spread Function (PSF)
\begin{equation}
    \text{PSF}(\mathbf{q}) \equiv I_{ref}(\mathbf{q}) \propto \left| \tilde{E}_\text{in}(\mathbf{q}) \ast \tilde{P}_{ref}(\mathbf{q}) \right|^2.
\end{equation}

By treating the unscattered central spot as a coherent guide-star, the refractive blurring can be computationally removed via standard Fourier deconvolution (e.g., employing a Wiener filter). The unblurred turbulence spectrum is retrieved as
\begin{equation}
    \langle |\tilde{V}(\mathbf{q})|^2 \rangle \propto \eta_0^2 \mathcal{F}^{-1} \left[ \frac{ \mathcal{F}\{I(\mathbf{q})\} }{ \mathcal{F}\{I_{ref}(\mathbf{q})\} } \right].
\end{equation}

This self-referencing property acts as a built-in computational adaptive optics mechanism. It mathematically guarantees that the Debye-Scherrer diffraction ring can be sharply reconstructed, and the turbulent kinetic energy directly measured even in near-critical plasma environments with severe density perturbations.



\section{Eddy Size Spectrum}

When transitioning the turbulent scattering model from a discrete, mono-disperse ensemble of eddies to a continuous, broad spectrum of sizes, different eddy sizes yield a continuous spectrum of diffraction angles $\vartheta \approx \lambda_L / l$). They would linearly superimpose to smear the Debye-Scherrer ring into a diffuse disk. This section shows that this polarimetric signal could be preserved because the diffraction from polarization drag is dominated by the large eddies. This differential diffraction acts as a spatial high-pass filter that forces the scattered intensity to peak sharply near the viscous dissipation scale.

For a continuous turbulent field, the ensemble-averaged turbulence form factor $\langle |\tilde{V}(\mathbf{q})|^2 \rangle$ transitions into the macroscopic spatial power spectrum of the turbulent vorticity field $S_\Omega(\mathbf{q})$. In incompressible turbulence, vorticity is the curl of the velocity field. In Fourier space, this spatial derivative introduces a factor of $q$, yielding the exact spectral relation for enstrophy
\begin{equation}
    S_\Omega(q) = 2 q^2 E(q)
\end{equation}
In a fully developed turbulent plasma, the kinetic energy follows the Kolmogorov power law in the inertial range, bounded by an exponential viscous cutoff at the dissipation scale $q_d = 2\pi/l_d$
\begin{equation}
    E(q) \approx C_K \varepsilon^{2/3} q^{-5/3} \exp\left[-\left(\frac{q}{q_d}\right)^2\right]
\end{equation}
where $C_K$ is the Kolmogorov constant and $\varepsilon$ is the turbulent dissipation rate. 

Substituting this kinetic energy spectrum into our polarimetric scattering relation, Eq.~(\ref{eq:Iperp}), yields the cross-polarized intensity distribution at the detector $\langle I(q) \rangle \propto q^{1/3} e^{-(q/q_d)^2}$, which reaches the peak intensity at $q_{peak} = q_d/\sqrt{6}$. Physically, the cross-polarized intensity $I(q)$ increases monotonically as $q^{1/3}$, suppressing the bright forward-scatter of the massive eddies  and leaving the interior of the diffraction pattern relatively dark. The intensity continues to pile up at larger scattering angles until it encounters the exponential viscous cutoff, forming a maximum at $q \approx 0.4 q_d$. 
Consequently, even when interrogating a continuous Kolmogorov cascade, the cross-polarized emission retains the annular halo structure similar to a Debye-Scherrer ring whose radius provides a measurement of the turbulence dissipation scale.

Actually, even if the plasma turbulence deviates from a standard Kolmogorov cascade, the kinetic energy spectrum $E(q)$ can be retrieved from the deconvolved detector images via $ E(q) \propto \frac{1}{q^2} \oint \langle |\tilde{V}(\mathbf{q})|^2 \rangle d\phi_q$. 
Because the cross-polarized scattering intensity inherently scales as $S_\Omega(q)\propto q^{2-p}$, the polarimetric diagnostic can successfully isolate and resolve any generalized turbulent energy cascade $E(q)\propto q^{-p}$ provided the spectrum is shallower than $p=2$, preventing the Scherrer-ring from defusing into a disk.
Thus, the radial intensity profile of the diffractive halo serves as a universal, assumption-free spectrometer for extracting the energy partition of any divergence-free plasma flow.
\section{Numerical Validation}

To illustrate the analytical model and the built-in phase deconvolution, we performed a 3D volumetric wave propagation simulation using a layered Grid-Monte Carlo phase-screen approach. The simulation tracks the simultaneous paraxial diffraction, nonlinear stochastic refraction, and polarization drag of an X-ray probe through a near-critical plasma ($n_0 = 0.99 n_c$) with macroscopic density fluctuations up to $0.004n_c$ (this number could be higher but it will demand a larger simulation plane to enclosure the diffracted beams). The underlying turbulence is prescribed by a Kolmogorov cascade bounded by a viscous cutoff at $l_d = 1\,\mu\text{m}$, simulated across $15$ longitudinal layers to capture cumulative volume scattering and phase distortions.

\begin{figure}[htbp]
    \centering
    \includegraphics[width=0.9\columnwidth]{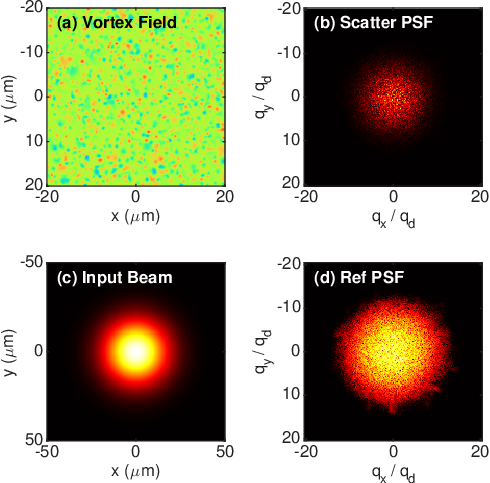} 
    \caption{(a) Vorticity field. 
    (b) The image of cross-polarized field intensity $I(\mathbf{q})$. 
    (c) The finite incident probe beam profile. 
    (d) The output main beam $\log_{10}I_{ref}(\mathbf{q})$, which suffers from stochastic refraction. }
    \label{fig:simulation}
\end{figure}

As demonstrated in Fig. \ref{fig:simulation}, the density fluctuation and the finite probe beam waist ($W_0 = 25\,\mu\text{m}$) distort both the incident beam and the cross-polarized scattered light, rendering the raw diffraction pattern unrecognizable. However, by utilizing the smeared incident beam as an instantaneous coherent guide-star, the Wiener deconvolution reverses the refractive distortion, as shown in Fig.~\ref{fig:simu}. It successfully recovers the isolated Debye-Scherrer ring, and exactly retrieves the theoretical $q^{-5/3}$ kinetic energy spectrum down to its dissipation limit.

\end{document}